\documentclass[a4paper]{jpconf}
\usepackage{graphicx}
\usepackage{amsmath}
\usepackage{amssymb}
\begin{document}
\title{HIKE: High Intensity Kaon Experiments at the CERN SPS}

\author{Matthew Moulson, for the HIKE Collaboration}

\address{INFN Laboratori Nazionali di Frascati, 00044 Frascati RM, Italy}

\ead{moulson@lnf.infn.it}

\begin{abstract}
  The availability of high intensity kaon beams at the CERN SPS North Area
  gives rise to unique possibilities for sensitive tests of the Standard Model
  in the quark flavor sector. Precise measurements of the branching ratios
  for the flavor-changing neutral current decays $K\to\pi\nu\bar{\nu}$ can
  provide unique constraints on CKM unitarity and, potentially, evidence for
  new physics. Building on the success of the NA62 experiment, plans are
  taking shape at CERN for a comprehensive program that will include
  experimental phases to measure the branching ratio for
  $K^+\to\pi^+\nu\bar{\nu}$ to $\sim$5\% and to $K_L\to\pi^0\nu\bar{\nu}$
  to $\sim$20\% precision. These planned experiments would also carry out
  lepton flavor universality tests, lepton number and flavor conservation
  tests, and perform other precision measurements in the kaon sector, as
  well as searches for exotic particles in kaon decays. We overview the
  physics goals, detector requirements, and project status for the next
  generation of kaon physics experiments at CERN.
\end{abstract}

\section{Introduction}
\begin{figure}[htbp]
\centering
\includegraphics[width=0.4\textwidth]{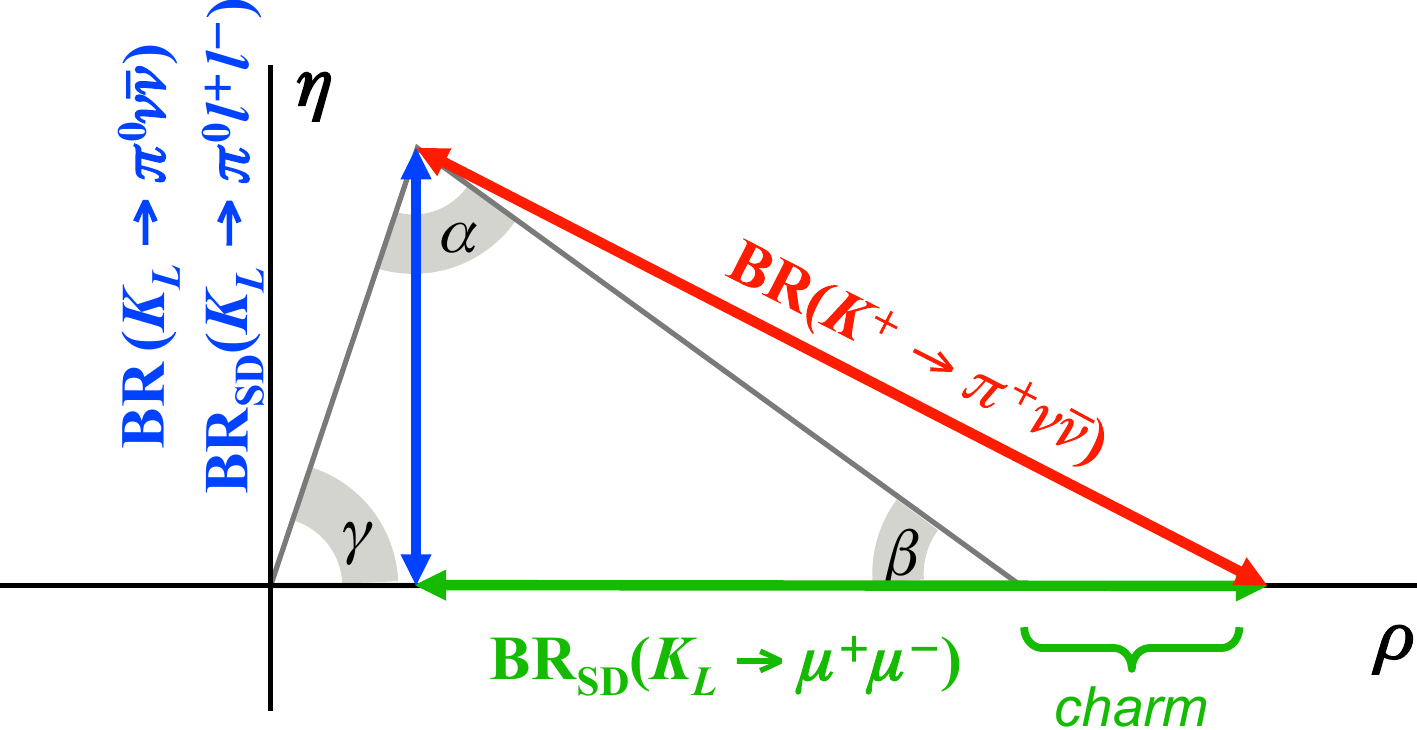}
\vspace{-2mm}
\caption{Determination of the unitary triangle with rare kaon decays.} 
\label{fig:ut}
\end{figure}

Rare kaon decays provide information on the unitary triangle, as illustrated
in Figure~\ref{fig:ut}. These are flavor-changing neutral current processes
(FCNC) that probe the $s\to d\nu\bar\nu$ or $s\to d\ell^+\ell^-$ transitions.
They are highly GIM-suppressed and their SM rates are very small.
The $K\to\pi\nu\bar\nu$ decays are the least affected by long-distance
physics. The branching ratios (BRs) for the $K\to\pi\nu\bar\nu$ decays are
among the observable quantities in the quark-flavor sector most sensitive
to new physics.

In the SM, the uncertainties on the $K\to\pi\nu\bar{\nu}$ rates are entirely
dominated by the uncertainties on the CKM matrix elements $|V_{ub}|$ and
$|V_{cb}|$ and the angle $\gamma$.
Using values for these parameters from the analysis of tree-level
observables, Buras et al. obtain~\cite{Buras:2015qea}
\begin{equation}
  \begin{split}   
    {\rm BR}(K_L\to\pi^0\nu\bar{\nu}) & = (3.4 \pm 0.6)\times10^{-11},\\
    {\rm BR}(K^+\to\pi^+\nu\bar{\nu}) & = (8.4 \pm 1.0)\times10^{-11}.
  \end{split}
  \label{eq:buras}
\end{equation}
Assuming no new-physics effects in $\epsilon_K$ and $\sin{2\beta}$ from
$B\to J/\psi K_S$, the BRs can be determined independently of $|V_{cb} |$ as
${\rm BR}(K_L\to\pi^0\nu\bar{\nu}) = (2.94 \pm 0.15)\times10^{-11}$ and
${\rm BR}(K^+\to\pi^+\nu\bar{\nu}) = (8.60 \pm 0.42)\times10^{-11}$~\cite{Buras:2021nns}.
The intrinsic theoretical uncertainties, if the CKM parameters are taken to
be exact, are about 1.5\% and 3.5\%, respectively.

Because the amplitude for $K^+\to\pi^+\nu\bar{\nu}$ has both real and imaginary
parts, while the amplitude for $K_L\to\pi^0\nu\bar\nu$ is purely imaginary, the
decays have different sensitivity to new sources of $CP$ violation. Measurements
of both BRs would therefore be extremely useful not only to uncover evidence of
new physics in the quark-flavor sector, but also to distinguish between new
physics models. More generally, measurement of all of the FCNC kaon decays
would allow the unitarity triangle to be overconstrained as illustrated in
Figure~\ref{fig:ut}, potentially providing evidence of new physics
independently of and in comparison to results from $B$ and $D$ meson decays
and providing important information about the flavor structure of that physics.

\section{The NA62 experiment}

NA62 is a fixed-target experiment at the CERN SPS, the goal of which is to
measure ${\rm BR}(K^+\to\pi^+\nu\bar{\nu})$ to within 10\%~\cite{NA62:2017rwk}.
The secondary positive beam, consisting of 6\% $K^+$ with a total rate of
750~MHz, is derived from the 400-GeV primary proton beam from the SPS
incident at zero angle on a beryllium target at a rate of $3\times10^{12}$
protons per pulse (ppp). 
The $K^+\to\pi^+\nu\bar{\nu}$ decay is detected in flight.
The signature is a $K^+$ entering the experiment and a $\pi^+$ leaving the
decay vertex, with missing momentum at the vertex and no other particles
observed in the final state. Principal backgrounds include those from the
abundant decays $K^+\to\mu^+\nu$ and $K^+\to\pi^+\pi^0$, as well as
backgrounds from upstream decays and interactions. The signal decay is
identified via selection in the $(p_\pi, m^2_{\rm miss})$ plane to exclude the
abundant two-body decays, where $p_\pi$ is the momentum of the candidate
pion track and $m^2_{\rm miss}$ is the squared missing mass at the vertex.
The distinguishing features of the experiment include high-rate, precision
tracking for both the beam and secondary particles, redundant particle
identification systems and muon vetoes, and hermetic photon vetoes, including
a high-performance EM calorimeter.

Between 2016 and 2018, NA62 observed more than $4\times10^{12}$ $K^+$ decays in
its fiducial volume, with the expectation of observing 10 signal events and 7
background events, principally from upstream decays and interactions. A total
of 20 events were observed, establishing the $K^+\to\pi^+\nu\bar{\nu}$ decay
with $3.4\sigma$ significance and providing the most precise measurement to
date for the branching ratio~\cite{NA62:2021zjw}:
\begin{displaymath}
  {\rm BR}(K^+\to\pi^+\nu\bar{\nu}) =
  (10.6^{+4.0}_{-3.4}|_{\rm stat}\pm0.9_{\rm syst})\times 10^{-11}.
\end{displaymath}

NA62 resumed data taking in July 2021 with a number of key modifications to
the beamline and detector to reduce background from upstream decays
and interactions and to allow data to be taken at the full nominal beam
intensity. The experiment is approved for data taking up to LHC Long Shutdown 3
(LS3), currently foreseen for the end of 2025, and is expected to measure
${\rm BR}(K^+\to\pi^+\nu\bar{\nu})$ to 10\% precision by then.
In the longer term, fixed-target runs in the SPS North Area are foreseen at
least through 2040. There is therefore an opportunity at the SPS for an
integrated program to pin down new physics in the kaon sector via measurement
of {\em all} rare kaon decay modes---both charged and neutral.

\section{The HIKE physics program}

The HIKE program (High Intensity Kaon Experiments at the
CERN SPS)~\cite{HIKE:2022aaa} is foreseen
to include three experimental phases for the comprehensive, high-precision
study of rare kaon decays during the period from the end of LS3 to the FCC era:
\begin{itemize}
\item {\bf Phase 1:} A $K^+$ experiment running at four times the intensity of
  NA62 ($1.2\times10^{13}$ ppp) to measure ${\rm BR}(K^+\to\pi^+\nu\bar{\nu})$
  to 5\% precision. Phase 1 will also focus on studies of lepton
  universality/number/flavor violation through observables such as
  $R_K = \Gamma (K^+\to e^+\nu)/\Gamma(K^+\to\mu^+\nu)$ and
  $K^+\to\pi^+\ell\ell$, and searches for decays such as 
  $K^+\to\pi^-\ell^+\ell^+$ and $K^+\to\pi^+\mu e$.
  The experiment will also make precision measurements of leptonic and
  semileptonic, radiative, and Dalitz decays and chiral parameters of the kaon
  system.
\item {\bf Phase 2:} An experiment with a neutral beam and tracking and PID
  systems optimized for the measurement of decays like
  $K_L\to\pi^0\ell^+\ell^-$ and $K_L\to\mu^+\mu^-$, as well as studies of lepton
  universality/number/flavor violation in $K_L$ decays,
  radiative $K_L$ decays and precision measurements, and measurements
  of $K_L$, $n$, and $\Lambda$ fluxes in the neutral beam and halo
  to prepare for the final phase.
\item {\bf Phase 3:} An experiment to measure
  ${\rm BR}(K_L\to\pi^0\nu\bar{\nu})$ to 20\%, also known as
  KLEVER~\cite{Ambrosino:2019qvz}.
\end{itemize}
In addition, periodic runs will be taken in beam-dump mode, with the target
out and the collimator in the secondary beam line closed, to allow sensitive
searches for the decays of exotic, long-lived particles produced in the decays
of mesons from interactions in the beam dump, with the goal of collecting
$10^{19}$ protons on target (pot) in dump mode in Phase 1 and up to
$5\times10^{19}$ pot by the end of Phase 3.

The experimental setup for all three phases will rely to the maximum extent
possible on the reuse of the same detectors in different configurations.
In particular, when a detector for HIKE is newly built or extensively
upgraded, if at all possible, it is designed to meet the performance
requirements for all successive phases.

\subsection{HIKE Phase 1}

\begin{figure}[htbp]
\centering
\includegraphics[width=\textwidth]{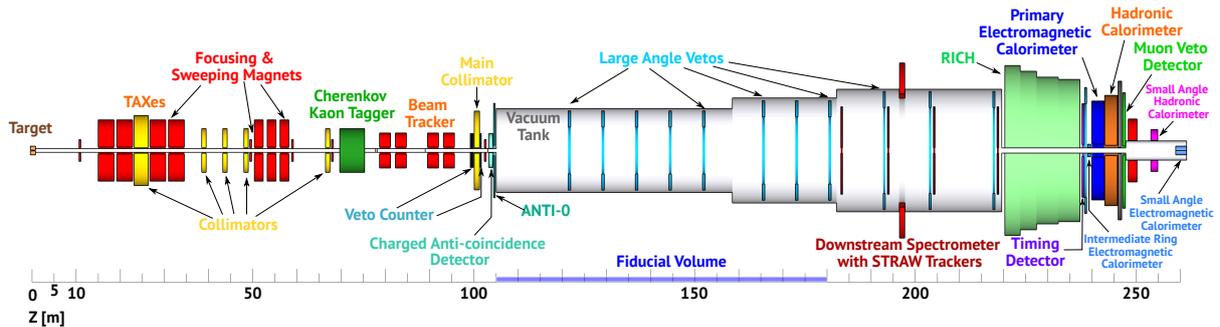}
\vspace{-2mm}
\caption{Experimental setup for HIKE Phase 1}
\label{fig:phase1}
\end{figure}
The four-fold increase in primary intensity needed for the measurement of
${\rm BR}(K^+\to\pi^+\nu\bar{\nu})$ to 5\% will require major upgrades of the
primary and secondary beamlines, as discussed in \cite{Gatignon:2650989}.
From the standpoint of the experiment, the success of NA62 has validated the
measurement technique and proves that the background can be handled.
The key challenge from the intensity increase is that the time resolution of
the detectors must be improved across the board by a factor of four in order
to maintain the loss of events from accidental coincidence (random veto) to
acceptable levels ($\lesssim25\%$), which must be achieved while maintaining
other key performance specifications such as space-time reconstruction
performance, low material budget, high single photon efficiencies, etc.
The experimental setup, shown in Figure~\ref{fig:phase1}, is not very different
from that of NA62, but most detectors will need to be rebuilt or extensively
overhauled to secure the needed performance.

Of particular interest, the
NA62 beam tracker (Gigatracker, GTK), consisting of three stations of silicon
pixel detectors, will need to be upgraded to track at 3~GHz. A time resolution
of better than 50~ps will be required, and the detector will have to be able
to handle rates of 8~MHz/mm$^2$ and be radiation resistant up for particle
fluences of more than $2\times10^{15}$ $n$~eq/cm$^2$/yr.
An excellent candidate technology is provided by the timeSPOT
project~\cite{hep-ph_Lai_2018,hep-ph_ignite_ALai_2022}, which is developing
hybrid 3D-trenched pixels in which the pixel electrode geometry is optimized
for timing performance.

The experiment's rate capability for secondary particles must be improved as
well. New straw-tube designs are being developed at CERN, based on past
collaboration with Dubna, 
that will allow straw chambers for use in vacuum to be developed
with 5-mm diameter straws with wall thickness of 20~$\mu$m.

It is natural to inquire whether the NA48 liquid-krypton (LKr)
calorimeter~\cite{Fanti:2007vi} used in NA62 can be reused for any of
the HIKE phases. NA48-era studies suggest and NA62 experience confirms that
the photon detection efficiency is sufficient for at least Phases 1 and 2.
The time resolution, however, is insufficient for the high-intensity program
and would require improvement by at least a factor of four.
Two directions are being pursued.
The first is to examine whether the LKr can be used for HIKE Phase 1.
In addition to necessary consolidation work, this would require upgrades to
make the calorimeter faster, including an increase in the operating voltage
to increase the drift velocity and faster signal shaping and digitizers for
the readout system.
For the $K_L$ phases, the diameter of the LKr inner bore limits the solid
angle of the beam that can be used, so a new calorimeter would be necessary
in any case. The new calorimeter could also be used in Phase 1, if it is ready
in time. An ideal choice of technologies appears to be the fine-sampling
shashlyk design used for the KOPIO and PANDA calorimeters~\cite{Atoian:2007up}, which has been shown to provide excellent energy and time resolution.
For HIKE, PID capability could be added by including independently read out,
1-cm-thick ``spy tiles'' at key points in the shashlyk stack (for example, at
the front of the calorimeter for charged-particle identification, at shower
maximum, and deep in the stack). Prototypes with this design have been tested
at Protvino and DESY. Another option under investigation is to construct the
calorimeter with new-generation, nanocomposite
scintillators~\cite{Gandini:2020aaa}, which offer high light output, fast
response, and good radiation robustness.

\subsection{HIKE Phase 2}

\begin{figure}[htbp]
\centering
\includegraphics[width=\textwidth]{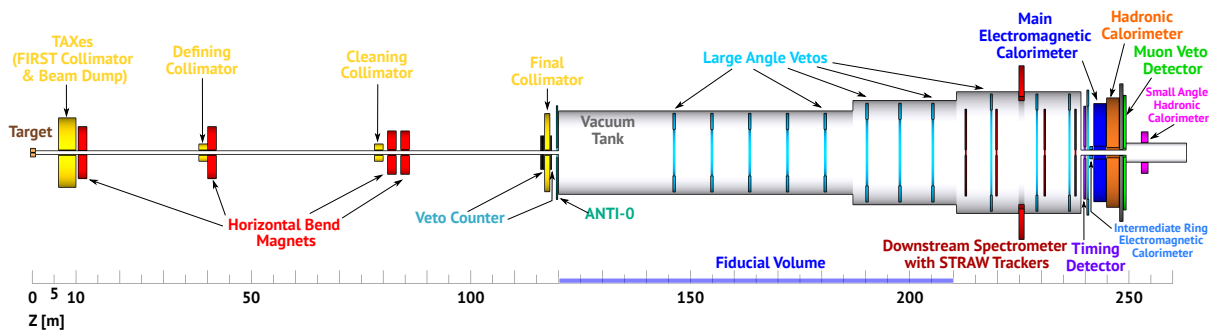}
\vspace{-2mm}
\caption{Experimental setup for HIKE Phase 2}
\label{fig:phase2}
\end{figure}
For HIKE Phase 2, a new neutral beamline is required. The baseline design is
the original 120-m beamline for KLEVER~\cite{VanDijk:2019oml}, featuring four
collimation stages, including an active final collimator that is incorporated
into the experiment and an oriented-crystal-metal photon converter at the
center of the dump collimator to reduce the flux of prompt photons in the
beam~\cite{Soldani:2022ekn}.
The collimation system defines a beam opening angle of 0.4~mrad.
The neutral beam is produced at $\theta=2.4$~mrad; $K_L$ mesons in the beam
have an average momentum of 79~GeV, while those decaying in the fiducial volume
(FV) have an average momentum of 46~GeV.
Relative to the Phase-1 experiment, in addition to the changes to the beamline
and the removal of the beamline detectors for charged particles, the RICH
is removed and the spectrometer is moved further downstream to increase the
acceptance for decays such as $K_L\to\pi^0\ell^+\ell^-$.
The beamline and experimental setup are illustrated in Figure~\ref{fig:phase2}.

At a primary intensity of $2\times10^{13}$~ppp (a six-fold increase with respect
to NA62), nearly $2\times10^{14}$ kaon decays will be observed in the FV in five
years of running. This will allow single-event sensitivities for
$K_L\to\pi^0\ell^+\ell^-$ to be improved by two orders of magnitude with
respect to the current best limits from
KTeV~\cite{AlaviHarati:2003mr,AlaviHarati:2000hs}.
There are topologically identical backgrounds to these channels from
$K_L\to\gamma\gamma\ell^+\ell^-$, with BRs that are 3--4 orders of magnitude
greater than for $K_L\to\pi^0\ell^+\ell^-$~\cite{Greenlee:1990qy}. 
Suppression of these backgrounds relies on the excellent energy
resolution of the HIKE EM calorimeter for the reconstruction of the $\pi^0$
mass peak for the signal decay.
HIKE Phase 2 will be well positioned to make the first observation of
the $K_L\to\pi^0\ell^+\ell^-$ decay, as well as to measure
${\rm BR}(K_L\to\mu^+\mu^-)$ with a statistical precision of 0.2\%, improving
on the current best result from BNL-E871~\cite{E871:2000wvm}, and should also
achieve BR sensitivities of $O(10^{-12})$ for a broad range of rare and
forbidden $K_L$ decays, such as lepton-flavor violating processes,
representing improvement on current best limits from BNL-E871 by up to a factor of 60.

\subsection{HIKE Phase 3}
\begin{figure}[htbp]
\centering
\includegraphics[width=\textwidth]{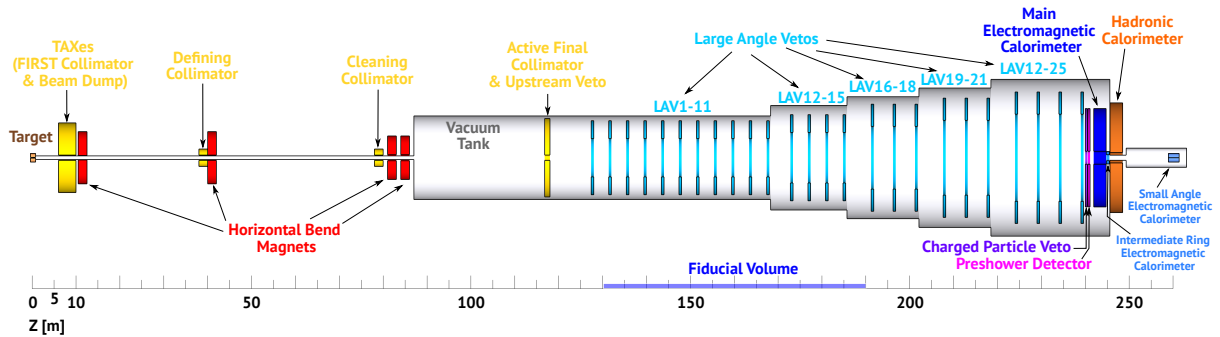}
\vspace{-2mm}
\caption{Experimental setup for HIKE Phase 3 (KLEVER)}
\label{fig:phase3}
\end{figure}

HIKE Phase 3 is a dedicated experiment, known as KLEVER, to measure
${\rm BR}(K_L\to\pi^0\nu\bar{\nu})$ to 20\%. Specifically, with a total
exposure of $6\times10^{19}$ pot in five years at an intensity of
$2\times10^{13}$ ppp, the KLEVER goal is to detect 60 signal events at
the Standard Model branching ratio, with a signal-to-background ratio
$S/B\sim 1$. The experiment is complementary to KOTO, in the sense that
the beam energy is significantly higher. As a result, photons from $K_L$
decays receive a significant boost, which makes photon vetoing easier.
On the other hand, the length of the experiment is much greater, and a
very long beamline is required to allow $\Lambda$ baryons and $K_S$ mesons
to decay upstream of the fiducial volume.
Relative to the Phase-2 beamline, the KLEVER beamline needs to be extended
by an additional 150~m from target to final collimator. This in turn requires
a downstream extension of the ECN3 hall by a similar amount.
For KLEVER running, the production angle for the neutral beam will be
increased from 2.4 to 8.0~mrad: this decreases the $n/K_L$ and
$\Lambda/K_L$ ratios for the beam and softens the momentum spectra so that
the average $K_L$ momentum is 39~GeV at production (26~GeV for $K_L$ mesons
that decay in the FV). This also decreases the absolute $K_L$ flux, which is
further reduced by the need to collimate the beam more tightly
($\Delta\theta=0.256$~mrad) for the extended beamline.
The need for additional construction to extend the ECN3 hall is a major
factor in scheduling KLEVER towards the end of the HIKE program.
Cost estimates are in progress.

From the standpoint of the experiment, the layout of the vacuum tank
and fiducial volume is roughly the same as for the other HIKE phases.
The spectrometer will be removed and the number of large-angle photon
veto detectors will be increased from 12 to 25 to extend the polar angle
coverage out to 100~mrad (from 50~mrad for Phases~1 and~2). The HIKE
large-angle vetoes themselves will be fine-segmented lead/scintillating
tile detectors similar to the VVS detectors for the planned but never
realized CKM experiment~\cite{Ramberg:2004en}. One particularly challenging
detector for KLEVER is the small-angle calorimeter (SAC), which sits in
the neutral beam at the downstream end of the experiment and which must
reject photons from background decays such as $K_L\to\pi^0\pi^0$ that
would otherwise escape via the beam exit. The SAC must have good photon
detection efficiency, especially for high energy photons (e.g., the
inefficiency must be $<10^{-4}$ for photons with $E>30$~GeV) while being as
insensitive as possible to the accidental coincidence of nearly 600~MHz of
neutral hadrons ($n$ and $K_L$) in the beam. The SAC must also have
$\sigma_t < 100$~ps, be able to separate two pulses a few ns apart, and be
radiation hard to $10^{14}$ $n$/cm$^2$ and $10^5$--$10^6$~Gy. Our proposed
solution is an ultra-fast, high-$Z$ crystal calorimeter based on a Cerenkov
radiator like PbF$_2$ or an ultra-fast scintillator such as
PWO-UF~\cite{Korzhik:2022xln}, which has a dominant emission component with
$\tau < 0.7$~ns.
The SAC will have transverse and longitudinal segmentation for $\gamma/n$
discrimination. From an engineering standpoint, it will be very similar to
the CRILIN calorimeter~\cite{Ceravolo:2022rag}, and R\&D is proceeding in
concert between the HIKE and CRILIN collaborations. An additional possibility
under investigation is to exploit the coherent interactions of high-energy
photons in oriented crystals to stimulate early pair conversion, allowing
the calorimeter to be realized with reduced thickness, increasing the
transparency to neutral hadrons while maintaining high detection efficiency
for photons~\cite{Bandiera:2019tmg}.

\section{Conclusions}

The HIKE project consists of a three-phase experimental program for the
comprehensive study of flavor physics in the kaon sector.
The experimental apparatus changes over time with a staged approach,
allowing HIKE to evolve and adapt its physics scope, an important feature
for a project that embraces a time scale of more than decade, during which
the physics landscape could change. Thanks to the successful experience of
NA62 and its predecessor NA48, the experimental techniques are well
established and robust expectations of sensitivity can be obtained from
the extrapolation of existing data. The HIKE Letter of Intent was submitted
to the CERN SPSC at the beginning of November 2022~\cite{HIKE:2022aaa}.
A formal proposal is in preparation for submission in fall 2023.
\vspace{2ex}

\section*{References}
\bibliographystyle{iopart-num}
\bibliography{Osaka_proc}

\end{document}